\documentclass[prl,twocolumn,showpacs,preprintnumbers]{revtex4}%
\usepackage{amsfonts}
\usepackage{amsmath}
\usepackage{amssymb}
\usepackage{subfigure}
\usepackage{graphicx}%
\begin{document}
\title[Magnetoresistance of BSCCO]{Bardeen-Stephen flux flow law disobeyed in the high-$T_c$ superconductor Bi$_{2}$Sr$_{2}$CaCu$_{2}$O$_{8+\delta}$}
\author{\'A.~Pallinger$^{1}$}
\author{B.~Sas$^{1}$}
\author{I.~Pethes$^{1}$}
\author{K.~Vad$^{2}$}
\author{F.~I.~B.Williams$^{1,3}$}
\author{G.~Kriza$^{1,4}$}
\affiliation{$^{1}$Research Institute for Solid State Physics and Optics, PO Box 49, H-1525
Budapest, Hungary}
\affiliation{$^{2}$Institute of Nuclear Research, PO Box 51, H-4001 Debrecen, Hungary}
\affiliation{$^{3}$CEA-Saclay, Service de Physique de l'Etat Condens\'{e}, Comissariat
\`{a} l'Energie Atomique, Saclay, F-91191 Gif-sur-Yvette, France}
\affiliation{$^{4}$Institute of Physics, Budapest University of Technology and Economics,
Budafoki \'ut 8, H-1111 Budapest, Hungary}

\pacs{74.72.Hs, 74.25.Fy, 74.25.Qt, 74.25.Sv}
\date{\today}

\begin{abstract}
Pulsed high current experiments in single crystals of the high-$T_{c}$
superconductor Bi$_{2}$Sr$_{2}$CaCu$_{2}$O$_{8+\delta}$ in $c$-axis directed 
magnetic field $H$ reveal that the $ab$-face resistance in the free flux flow regime is a 
solely logarithmic function of H, devoid 
of any power law component. Re-analysis of published data confirms this result 
and leads to empirical analytic forms for the $ab$-plane and $c$-axis
resistivities: $\rho_{ab}\propto$ $H^{3/4}$, which does not obey the expected
Bardeen-Stephen result for free flux flow, and $\rho_{c}\propto H^{-3/4}%
\log^{2}H.$
\end{abstract}
\maketitle

Free flux flow (FFF) resistivity describes 
how fast the vortices in a type II superconductor move in the
direction of an applied force \cite{kopnin}. It is a measure of how the
momentum of the superfluid is transferred to the host lattice via
quasiparticle excitations. The
velocity-force relation expressed by the FFF resistivity has to be taken into
account in interpreting any vortex transport, be it global or local. 
Since the primary source of dissipation in a type II superconductor in magnetic
field is vortex motion, FFF resistivity is also of great importance for 
technical applications. 

A transport current exerts a Lorentz-Magnus force on a vortex and if 
other forces like vortex-defect interaction (pinning) are negligible (i.e., the 
vortex motion is ``free''), the velocity-force relation can be inferred from the 
resistivity $\rho_{\mathrm{FFF}}$. The Bardeen-Stephen (BS) law \cite{bardeen1965} 
states that $\rho_{\mathrm{FFF}}$ is proportional to the density of vortices and 
therefore to the magnetic field $H$:
\begin{equation}
\rho_{\mathrm{FFF}}=\gamma\rho_{n}(H/H_{c2})^{\beta},\ \ \beta=1, \label{eq:bs}%
\end{equation}
where $\rho_{n}$ is the normal state resistivity, $H_{c2}$ the upper critical
field, and $\gamma$ a constant $\approx 1$. This law has been
experimentally established \cite{parks} for a number of conventional
superconductors. In high-$T_{c}$ materials the quasi-two-dimensional (2d) 
electronic structure, the nodes of the $d$-wave order parameter, and the
structure of the vortex system may potentially influence FFF.
The motion of 2d ``pancake'' vortices along their well conducting $ab$ plane 
leads to dissipation by flux flow, whereas in the poorly conducting 
$c$ direction dissipation is governed not by flux flow but by tunneling 
between weakly coupled $ab$ planes. 
A quasiclassical calculation \cite{kopnin1997} suggests 
that the BS law is valid also in the $d$-wave case. 

Experiments to 
test the validity of Eq.\ (\ref{eq:bs}) in
high-$T_{c}$ superconductors---and especially in Bi$_{2}$Sr$_{2}$CaCu$_{2}%
$O$_{8+\delta}$ (BSCCO), the model system of this study---are contradictory. 
Data on low frequency transport in single crystals
\cite{bs-bscco} and thin films \cite{thinfilm,xiao} as well as microwave and
millimeter wave impedance \cite{microwave} are inconsistent with one another,
agreeing only that \textit{the BS law is not obeyed.} 
The resistivity often resembles a sublinear power law in field. The situation is 
similar in other high-$T_{c}$ materials with the notable
exception of the results of Kunchur \textit{et al.}\ \cite{kunchur} who find
agreement with BS law in thin film resistivity measurements in 
YBa$_{2}$Cu$_{3}$O$_{7}$. However, the resistance they measure 
does not saturate, i.e., it increases with current, up to the highest current they use, 
leading to an uncertainty in the value of $\rho_\mathrm{FFF}$. A re-analysis of the 
differential resistance indicates again a sublinear field dependence of $\rho_\mathrm{FFF}$.

The main difficulty in measuring the velocity-force relation is to take
account of the pinning force about which one has little detail. One way is to model 
pinning to interpret the surface impedance
arising from local vortex motion. Another approach is
to create experimental conditions where pinning is irrelevant as occurs in a
true (unpinned) vortex liquid. We extend our experiments to the non-ohmic regime
by applying sufficiently high current that the pinning force is negligible
compared with the Lorentz-Magnus force from the transport current. 

To this end we have made pulsed high-current transport measurements on single
crystal BSCCO with electrode contacts on the face parallel to the well conducting 
$ab$ planes in a $c$-axis directed magnetic field.
The \textit{global} single crystal resistance measured on the $ab$-face 
in the ohmic regime and 
the asymptotic high-current differential resistance in the non-ohmic regime
show the same logarithmic magnetic field dependence
devoid of any power law. We combine this result with published data 
from other experiments \cite{busch,morozov} and set up  
empirical functional forms for the \textit{local} resistivities 
$\rho_{ab}$ and $\rho_{c}$, valid over a broad range of temperature and field 
in the vortex liquid phase. 
The most striking and important of our conclusions is
that for BSCCO $\beta=3/4$ in Eq.\ (1).

We selected for experiment three single crystals from 3 different batches of Bi$_{2}$Sr$_{2}$CaCu$_{2}%
$O$_{8+\delta}$ with typical dimensions $1\times0.5\times0.003$
mm$^{3}$, the shortest corresponding to the poorly conducting $c$ axis.
All were close to optimal doping with a resistance-determined critical
temperature $T_{c}\approx 89$~K and transition width about $2$~K in zero field; 
the diamagnetism in a 1 mT field set in
progressively below $T_{c}$ to near $100\%$ at low temperature. 

Voltage-current ($V$-$I$) response was measured in the usual 
four-point configuration on an $ab$ face
in perpendicular magnetic field with two current contacts across the width
near the ends and two point voltage contacts near each edge of the same face.
The contacts were made by bonding 25~$\mu$m gold wires with silver epoxy fired
at $900$~K in an oxygen atmosphere resulting in current contact resistances of
less than 3~$\Omega$. To avoid significant Joule heating, we employed short
($\leq50$ $\mu$s) current pulses of isosceles triangular shape at $0.2$ to $1$
s intervals. Technical details and the issue of Joule heating are treated in
Ref.\ \cite{sas2000} with the conclusion that the temperature change in the
area between the voltage contacts is negligible for the duration of the pulse.

\begin{figure}[ptb]
\includegraphics[width=6.5cm]{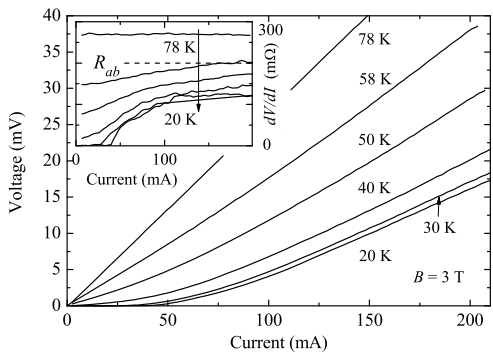}\caption{ Typical voltage-current
characteristics at selected temperatures in $B=3$ T. Inset: current dependence of differential
resistance $dV/dI$ at the same temperatures. At high currents $dV/dI$ saturates at $R_{ab}$.}%
\label{fig:iv}%
\end{figure}

\begin{figure}[ptb]
\includegraphics[width=8.4cm]{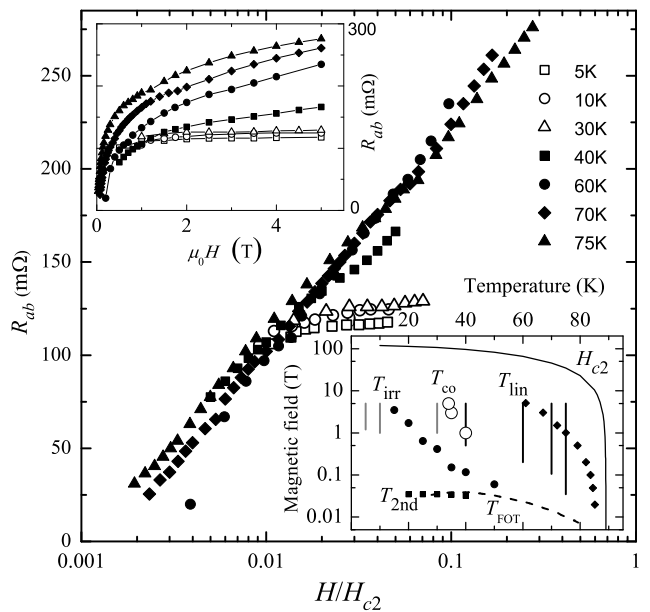}
\caption{
High-current differential
resistance $R_{ab}$ at several temperatures as a function of the logarithm of magnetic field 
normalized to the upper critical field $H_{c2}(T)$. Upper inset: Same
data on a linear field scale. Lower inset: 
Phase diagram measured on same samples except for the first order transition $T_{\mathrm{FOT}}$
taken from Ref.\ \cite{zeldov} for comparison. $T_{\mathrm{irr}}$ refers to 
magnetic irreversibility, $T_{\mathrm{2nd}}$ to second magnetization peak, 
$T_{\mathrm{lin}}$ to the beginning of linear $V$-$I$.
$H_{c2}=120 [1-(T/T_{c})^{2}]$ tesla. 
The vertical black (grey) lines show the range of full (open) symbols in the main panel.
The open circle is crossover in the field dependence of $R_{ab}$.
}%
\label{fig:2}%
\end{figure}

Typical $V$-$I$ characteristics at different temperatures in a field of $3$~T
are shown in Fig.\ \ref{fig:iv}. Above a temperature $T_{\mathrm{lin}}<T_c$ the $V$-$I$ 
curves are linear (see the lower inset of Fig.\ \ref{fig:2} for the field dependence of the
characteristic temperatures).
Below $T_{\mathrm{lin}}$ nonlinearity develops and the $I\rightarrow 0$ resistance decreases 
faster than exponentially until it becomes unmeasurably small even with the most sensitive technique. 
At low temperature dissipation
sets in abruptly at a threshold current $I_{\mathrm{th}}$; for higher temperatures a marked upturn 
in the $V$-$I$ curve (a ``knee'') is seen at $I_{\mathrm{k}} \lesssim I_{\mathrm{th}}$. 
Throughout the nonlinear range 
the differential resistance increases with increasing current; 
for currents several times $I_{\mathrm{th}}$ or $I_{\mathrm{k}}$ it saturates 
(becomes current independent) at a value $dV/dI=R_{ab}$ as shown in the 
inset of Fig.\ \ref{fig:iv}. Since $I_{\mathrm{th}}$
and $I_{\mathrm{k}}$ are hallmarks of depinning, 
the current-independent $R_{ab}$ observed at many times this current suggests 
that at these high currents pinning is irrelevant and $R_{ab}$ reflects 
FFF \cite{nonlin}. The focus of this article is the behavior of $R_{ab}$. 

The upper inset of Fig.~\ref{fig:2} shows the field dependence of $R_{ab}$ 
for the temperatures indicated in the
phase diagram of the lower inset. The resistance is field and temperature 
independent at low temperature. 
With increasing temperature there is 
a crossover to a field-dependent behavior
at $T_{\mathrm{co}}$ (open circles in the phase diagram) 
situated between the magnetic irreversibility line  and $T_{\mathrm{lin}}$. 
Having in mind that in the BS law the characteristic
field is $H_{c2}$, we interpolate the upper critical field using the form
$H_{c2}=(120\ \mathrm{tesla})[1-(T/T_{c})^{2}]$ constructed from
$dH_{c2}/dT_{T_{c}}=-2.7$ T/K \cite{li1993} and use it to plot $R_{ab}$
against $H/H_{c2}$ on a logarithmic field scale in the main panel of Fig.\ \ref{fig:2}.
The high-temperature curves all collapse into a master curve representing a
logarithmic field dependence:
\begin{equation}
R_{ab}(H,T)=R_{ab}^{n}[1+\alpha\log(H/H_{c2}(T))]\ ,
\label{eq:Rf}%
\end{equation}
where $R_{ab}^{n}$ is the zero field normal resistance at $T_{c}$ 
and $\alpha$ is a constant. This scaling only contains the
temperature through $H_{c2}(T)$. 
Equation (\ref{eq:Rf}) provides an excellent description of all three samples; 
for the parameter $\alpha$
we find 0.16, 0.19 and 0.21, essentially the same values, insensitive to
the presumably different disorder between samples. 

We emphasize that $R_{ab}$ cannot be compared directly to the BS law because
the strong anisotropy of the electronic properties makes the current
distribution very inhomogeneous and $R_{ab}$ reflects both $ab$-plane and
$c$-axis properties. However, if the sample is thick in the $c$ direction, a
simple scaling argument for a current independent local resistivity tensor
yields $R_{ab}=A\sqrt{\rho_{ab}\rho_{c}}$ where $A$ is a geometrical factor. 
This
relation is valid in the linear region $T>T_{\mathrm{lin}}(B)$ but also 
below $T_{\mathrm{lin}}$ if the current is sufficiently high that 
the current density is well in
the upper differentially linear portion of the response near the top surface 
of the crystal \cite{nonlin}. 
The analysis and experimental checks of
Ref.\ \cite{busch} indicate that with the sample size, shape, and contact
geometry used in their and our single crystal $ab$ plane studies in BSCCO, 
the thick sample limit provides a good description. 

\begin{figure}[ptb]
\includegraphics[width=7cm]{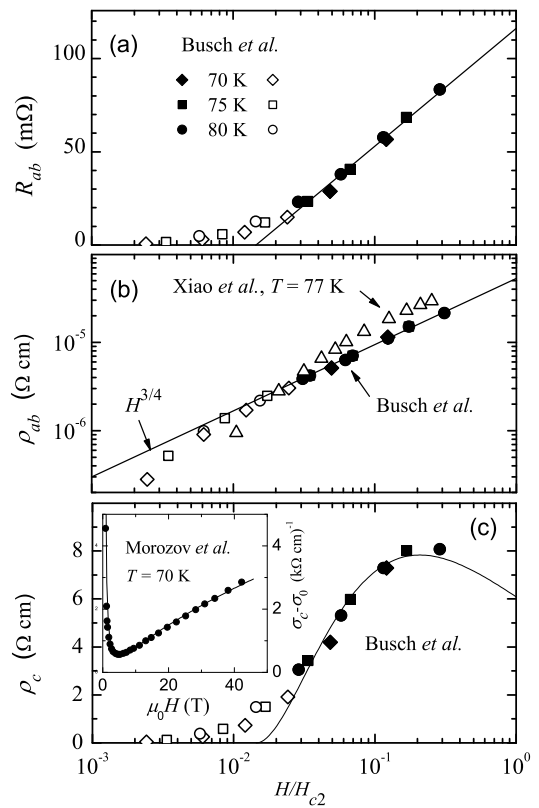}
\caption{ Single crystal resistance
$R_{ab}$ (a); $ab$ plane resistivity $\rho_{ab}$ (b); and $c$-axis resistivity
$\rho_{c}$ (c) as a fuction of magnetic field normalized to the upper critical
field $H_{c2}$. Data from Ref.\ \cite{busch}. For full symbols $T>T_{\mathrm{lin}}$.
Solid lines in panels (a) to (c) are fits to Eqs.\ (2) to (4), respectively. 
In panel (b) 77-K thin film data from Ref.\ \cite{xiao} is
also shown by triangles. Inset of panel (c): $c$-axis conductivity data 
from Ref.\ \cite{morozov}. $\sigma_{c}
=1/\rho_{c}$ and $\sigma_{0}$ a constant. The solid line is a fit to Eq.\ (3).
}%
\label{fig:3}%
\end{figure}

Independent confirmation of our results emerges from analysis of other
experiments. Although high-current data are absent, the fact that the $V$-$I$
curves are linear for $T>T_{\mathrm{lin}}(H)$ allows comparison with low-current data
in this temperature and field range. In Fig.\ \ref{fig:3}(a) we show $R_{ab}$
\textit{vs.}\ $H$ curves extracted from the $R_{ab}$ \textit{vs.}\ $T$ data
taken by Busch \textit{et al.}\ \cite{busch} in different magnetic fields.
Excellent agreement with Eq.\ (\ref{eq:Rf}) is seen for $T>T_{\mathrm{lin}}$ (full
symbols in the figure) with $\alpha=0.23$.

Busch \textit{et al.}\ \cite{busch} were able to disentangle $\rho_{ab}$ and
$\rho_{c}$ by using data from two additional contacts on the bottom of
the crystal. $\rho_{ab}$ and $\rho_{c}$ results extracted from their work for
a series of temperatures are shown as a function of $H/H_{c2}$ in Fig.\ \ref{fig:3}(b)
and (c). In the temperature and field range $T>T_{\mathrm{lin}}(H)$ both quantities
individually exhibit $H/H_{c2}$ scaling. The in-plane resistivity does
\textit{not} agree with the $\beta=1$ BS law of Eq.\ (1), but is well
described by a $\beta=3/4$ exponent (best fit $\beta=0.75\pm0.02$). 
Having definite analytic forms for both $R_{ab}$ and $\rho_{ab}$, we can use
the relation $(R_{ab}/A)^{2}=\rho_{ab}\rho_{c}$ to write an expression for the
$c$-axis resistivity. In summary:
\begin{eqnarray}
\rho_{ab}&\equiv & \rho_\mathrm{FFF}=\rho_{ab}^{n}(H/H_{c2})^{\beta}, \label{eq:rhoab}
\\
\rho_{c}&=&  \rho_{c}^{n}(H/H_{c2})^{-\beta}[1+\alpha\log(H/H_{c2})]^{2}, \label{eq:rhoc}
\\ 
\beta & = & 3/4,~\alpha=0.2~. \nonumber
\end{eqnarray}
The prefactors $\rho_{ab}^{n}$ and $\rho_{c}^{n}$ are in good agreement 
with the respective normal resistivities at $T_{c}$.

Are these forms corroborated by other types of measurement? In principle
$\rho_{ab}$ can be measured in thin films where the current density is
expected to be homogeneous. In Fig.\ \ref{fig:3}(b) we show the $77$ K
thin film resistivity obtained by digitizing the $V$-$I$ curves in
Ref.\ \cite{xiao}. Data above about 1 T are reasonably well described by a
$H^{3/4}$ dependence, but a closer look reveals that the $\log\rho_{ab}$
\textit{vs.}\ $\log H$ curves are concave from below at every $H$, i.e., there
is a systematic deviation from power law. This ``logarithm
like'' (but not logarithmic) dependence is shared with other
thin film results \cite{thinfilm} but there are significant
quantitative differences between data measured by different groups. A possible
reason is that macroscopic defects like steps on the surface or mosaic
boundaries force $c$-axis currents and the measured resistance is a
sample-dependent combination of $\rho_{ab}$ and $\rho_{c}$. 

The expression for $\rho_{c}$ reproduces well the maximum 
(at $H_{\mathrm{max}}=H_{c2}\exp(8/3-1/\alpha)\sim0.1H_{c2}$) 
observed in the high field $c$-axis
magnetoresistance \cite{morozov} and the overall field dependence of these
independently measured data is very well described by Eq.~(\ref{eq:rhoc}).
We demostrate this in Fig.\ \ref{fig:3}(c) where we plot $70$ K data for $\sigma
_{c}(B)-\sigma_{0}$ from Fig.\ 5 of Ref.\ \cite{morozov} where $\sigma
_{c}=1/\rho_{c}$ and the constant $\sigma_{0}$ is interpreted as the
zero-field quasiparticle conductivity. Using our estimate of $H_{c2}%
(T)\approx46$ T for $T=70$ K, Eq.\ (\ref{eq:rhoc}) fits the measured data with
parameters $\sigma_{c}^{n}=1/\rho_{c}^{n}=6.6$~(k$\Omega$cm)$^{-1}$,
$\sigma_{0}=3.6$~(k$\Omega$cm)$^{-1}$ and $\alpha=0.20$ to obtain the curve
indicated by the continuous line in the figure. The value of $\alpha$
is in excellent agreement with that inferred from $R_{ab}$ measurements.
It should be pointed out, however, that $\sigma_{0}$ is significantly smaller than the
value $\sigma_{0}\approx 8$~(k$\Omega$cm)$^{-1}$ inferred in
Ref.\ \cite{morozov}. In terms of resistivities, this means that $\rho_{c}$
decreases more slowly in high fields than described by Eq.\ (\ref{eq:rhoc}) with
$\beta=3/4$ and is in fact best described with an exponent $\beta=0.51$.

In the high-current limit the same form for $R_{ab}(H)$ also holds below
$T_{\mathrm{lin}}(H)$ where the $V$-$I$ curves are nonlinear. 
Since no
change in the behavior of $R_{ab}$ is observed when the $T_{\mathrm{lin}}(H)$ line is
crossed, it is reasonable to assume the same for $\rho_{ab}$ and $\rho_{c}$. 
In the low field direction a lower limit of the validity of Eq.~(\ref{eq:Rf}) 
is the zero of the equation at $H_0/H_{c2}=e^{-1/\alpha}\sim 10^{-3}-10^{-2}$, higher but in the order
of the first order transition in the static vortex system. In the high-field direction
Eq.~(\ref{eq:Rf}) is valid up to the highest field $\approx 0.3 H_{c2}$ we investigated.

The temperature $T_\mathrm{co}$ of the crossover from $R_{ab}=\mathrm{const}$ 
to $R_{ab}\propto \log H$ is distinctly higher than 
the onset of magnetic irreversibility at $T_\mathrm{irr}$, and also above 
the vortex glass transition $T_g\approx T_\mathrm{irr}$ inferred 
from scaling analysis \cite{glass} of the $V$-$I$ curves. 
On the other hand, no change in the behavior of $R_{ab}$ is 
observed when the $T_\mathrm{irr}(H)$ and $T_g(H)$ lines are crossed. 
This suggests that because the pinning potential is smoothed at high velocities, 
the phase diagram \cite{phasediag} of
the far-from-equilibrium \textit{dynamic} vortex system  \cite{dynvortex} 
is different from that of the unperturbed thermodynamic phases. 
Since $R_{ab}$ behaves the same in the pinned ($T<T_\mathrm{lin}$) and unpinned ($T>T_\mathrm{lin}$)
liquid phases, we propose that the unpinned phase, otherwise observed only above $T_\mathrm{lin}$, may be restored in the range 
$T_\mathrm{co} < T < T_\mathrm{lin}$.
Then $T_\mathrm{co}$ may approximate the melting transition in a hypothetical defect-free crystal.

Our most robust finding, invariably observed not only in our 3 batches but also in the 
data of Ref.~\cite{busch}, is the logarithmic field dependence of the high-current single
crystal resistance $R_{ab}$. Although a power of $H$ factor is expected both
in $\rho_{ab}$ \cite{kopnin} and $\rho_{c}$ \cite{vekhter}, no such factor is
present in $R_{ab}\propto\sqrt{\rho_{ab}\rho_{c}}$. The most likely reason is
that the power-law factors in $\rho_{ab}$ and $\rho_{c}$ cancel
(exponents 3/4 and -3/4 in our analysis). The cancellation is very
accurate; we estimate that a power law factor with exponent as low as 0.1
could be observed in our $R_{ab}$ data.
Moreover, because we find no
logarithmic correction to $\rho_{ab}$, the logarithmic dependence of
$R_{ab}$ is carried by $\rho_{c}$. 

Arguing that both $\rho_{ab}$ and
$\sigma_{c}$ are proportional to the quasiparticle density of states at the
Fermi level, $N(0)$, it cancels in the product $\rho_{ab}
\rho_{c}$. In conventional superconductors $N(0)$ is proportional to the
number of vortices therefore to $H$, leading to the $H$-linear resistivity of
the BS law. In nodal gap superconductors near-nodal quasiparticles lead to a
sublinear dependence; for line nodes $N(0)\propto H^{1/2}$
\cite{volovik}, as evidenced in recent low-temperature thermodynamic
measurements \cite{thermo}. Although delocalized 
near-nodal quasiparticles are not expected to contribute significantly to 
$\rho_\mathrm{FFF}$ because of the weak spectral flow force \cite{kopnin1997} 
they experience, 
the result may be different in the diffusive limit in the liquid phase. 
A possible reason for the $\beta=3/4$ exponent is the 
different structure factors of the solid and liquid phases.

Nonlocal effects \cite{levin,koshelev} may influence the evaluation of
the 6-contact measurements \cite{busch} and therefore the validity of 
Eqs.~(\ref{eq:rhoab}) and (\ref{eq:rhoc}) (but not of Eq.~(\ref{eq:Rf})). 
This seems, however, unlikely in the light of the good agreement of $\rho_c$ inferred from 
independent $ab$-plane \cite{busch} and $c$-axis \cite{morozov} measurements
and of the broad temperature range of validity of Eq.~(\ref{eq:Rf}). 

In conclusion, we have set up empirical rules for the analytic form of single
crystal resistance as well as for the $ab$ plane and $c$ axis 
resistivities in
the high-current free flux flow limit in the vortex liquid state of BSCCO,
valid over a broad range of temperature and field. Both the logarithmic field 
dependence of the single crystal resistance and the 3/4-power law in the 
$ab$-plane free flux flow resistance are in disagreement with the current
theoretical understanding of high-$T_c$ superconductors.  
\bigskip

We acknowledge with pleasure fruitful discussions with F.~Portier,
I.~T\"utt\H{o}, L.~Forr\'o and T.~Feh\'er and the help and technical expertise
of F.~T\'oth. L.~Forr\'o and the EPFL laboratory in Lausanne have contributed
in a very essential way to sample preparation and characterization. Finally we
acknowledge with gratitude the Hungarian funding agency OTKA (grant no.\ K 62866).

\bibliography{AnMR}

\begin{thebibliography}{23}
\bibitem[{non()}]{nonlin}
\bibinfo{note}{In a numerical simulation with nonlinear local conductivity we
  find that at $I>3I_\mathrm{k}$, $R_{ab}=dV/dI$ approximates the high-current
  linear resistance within a few percent. $dV/dI$ is always closer to the
  linear resistance than $V/I$.}
\end{thebibliography}

\end{document}